\documentclass[9pt,twocolumn,twoside]{osajnl}
\journal{ol} % Choose journal (ao, aop, josaa, josab, ol, pr)
\usepackage{graphicx}% Include figure files
\usepackage{dcolumn}% Align table columns on decimal point
\usepackage{lipsum}
\usepackage{bm}% bold math
\usepackage{tabu}
\usepackage[utf8]{inputenc}
\usepackage[T1]{fontenc}

\usepackage{color}

\usepackage{comment}
\usepackage{soul}
\usepackage{siunitx}

\setboolean{shortarticle}{true}
\title{Vectorial Helico-Conical Beams}

\author[1]{Edgar Medina-Segura}
\author[1]{Leonardo Miranda-Culin}
\author[2]{Valeria Rodríguez-Fajardo}
\author[3]{Benjamin Perez-Garcia}
\author[1*]{Carmelo Rosales-Guzm\'an}
\affil[1]{Centro de Investigaciones en Óptica, A.C., Loma del Bosque 115, Colonia Lomas del campestre, 37150 León, Gto., Mexico.}
\affil[2]{Department of Physics and Astronomy, Colgate University, Hamilton, NY 13346, United States of America.}
\affil[3]{Photonics and Mathematical Optics Group, Tecnologico de Monterrey, Monterrey 64849, Mexico.}
\affil[*]{Corresponding author: carmelorosalesg@cio.mx}
\begin{abstract} 
In this work, we propose and demonstrate experimentally a new family of vector beams, the Helico-Conical Vector Beams (HCVB), whose spatial degree of freedom is encoded in the Helico-Conical Optical Beams. We use Stokes polarimetry to study their properties and find that upon propagation their transverse polarisation distribution evolves from non-homogeneous to quasi-homogeneous, such that even though their global degree of nonseparability remains constant, locally it decreases to a minimum value as $z\rightarrow\infty$. We corroborated this quantitatively using the Hellinger distance, a novel metric for vectorness that applies to spatially-disjoint vector modes. To the best of our knowledge, HCVBs are the second family of vector beams in which this behaviour has been observed, paving the way for applications in optical tweezing or information encryption.
\end{abstract}

\setboolean{displaycopyright}{true}

\begin{document}
\maketitle
Since their inception in 2005, Helico-Conical Beams (HCB) have attracted the attention of the structured light community and  found several applications \cite{amadeo2005}. The HCB are generated from a phase function with helical and conical phase dependence, which results from the product between the radial and azimuthal coordinates $r$ and $\theta$, respectively. The first experimental realisation of such beams ignited an intense study of their many properties. For example, in 2007, the phase structure of the HCB was examined by Hermosa \textit{et. al.} and they concluded that \textit{``HCB in free-space propagation can be considered as superposition of integer charged vortices in a host beam"} \cite{HERMOSA2007178}. Five years later, in 2013, it was demonstrated that when an opaque obstacle obstructs HCBs, they experience a propagation-dependent self-healing effect \cite{Hermosa:13}. Additional studies include the propagation dynamics of HCB through scalar diffraction theory \cite{BAREZA2015236}, as well as the detection of the Orbital Angular Momentum (OAM) contained in such beams through interferometric techniques \cite{ENGAY2019247}. Variations of the original Helico-Conical phase have provided novel types of beams, which include the scalar superposition of multiple HCB \cite{alonzo2007three}, the generalisation of the helico-conical phase in the radial coordinate, which led to the Power-Exponent Helico-Conical Optical Beams (PEHCB) \cite{CHENG2019288}, the Modified Helico-Conical Beams (MHCB), which is a generalisation along the angular coordinate \cite{XIA2020124824} or the Bored Helico-Conical Beams \cite{Zeng:22}. On the application side, HCB have been used in optical tweezers \cite{opticalTwisters}, as well as in the fabrication of three-dimensional chiral microestructures \cite{wen2022fabrication}.

Almost in parallel, the field of complex vector beams has gained popularity, in part due to their wide variety of potential applications \cite{Rosales2018Review,Shen2022,Yuanjietweezers2021,Zhan2009,Ndagano2018,Toppel2014,BergJohansen2015,Fang2021}. Vector beams are general states of light featuring a non separability between their spatial and polarisation degrees of freedom, which give rise to light beams with all sorts of exotic nonhomogeneous transverse polarisation distribution. Moreover, even though the polarisation degree of freedom is bounded to a two-dimensional space, the spatial degree of freedom has infinite more options, giving rise to infinite variety of vector modes \cite{Dudley2013,rosales2021Mathieu,Zhao2022,Yao-Li2020,Galvez2012,Hu2021}. Along this line, the study of vector beams with a propagation-dependent state of polarisation is gaining popularity, not only due to its fundamental properties but also because it provides an alternative tool for the development of novel applications. Examples include vector beams whose polarisation distribution varies with propagation, which has been achieved with Bessel-  \cite{Moreno2015, ShiyaoFu2016, PengLi2017, PengLi2016}, Laguerre- \cite{Otte2018,zhong2021,PengLi2018} and self-focusing circular Airy-Gauss beams \cite{Hu2023,Hu2022abruptly}. The use of the Gouy effect has also given rise to vector beams that oscillate in a sinusoidal way between different vector beams \cite{zhong2021}. More intricate examples include vector beams where also the topological charge of the constituting scalar beams changes upon propagation \cite{Davis2016}, as well as those whose transverse polarisation structure can not only rotate but also accelerate \cite{Lu2020,Singh2021}. In a slightly different case, the vector beam reported in \cite{Hu2021} splits upon propagation along the transverse plane, evolving from a single beam to two spatially-disjoint beams of orthogonal polarisation. As result, its transverse polarisation also evolves from a mixture of linearly-polarised states to a quasi-homogeneous polarisation state, thus experiencing a decrease in their local degree of concurrence. This unusual behaviour inspired a novel technique to quantify the propagation-dependent inhomogeneity of spatially-disjoint vector beams \cite{Aiello2022}. 

In this work, we demonstrate theoretically and experimentally a new class of vector beams which we term  Helico-Conical Vector Beams (HCVB).  These beams are generated as the weighted superposition of two orthogonal circular polarisation states and two scalar HCB. By construction, at the generation plane ($z=0$) the HCVB features a non-homogeneous polarisation distribution that evolves upon propagation into a quasi-spatially disjoint spatial mode with quasi-homogeneous polarisation distribution. As a result, the local degree of concurrence of HCVB decreases as a function of propagation. To analyse this behaviour, we reconstructed the polarisation distribution at various planes using Stokes polarimetry and quantified the evolution from $z=0$ to $z\to\infty$, using the Hellinger distance proposed in \cite{Aiello2022}.

\begin{figure}[tb]
    \centering    \includegraphics[width=0.48\textwidth]{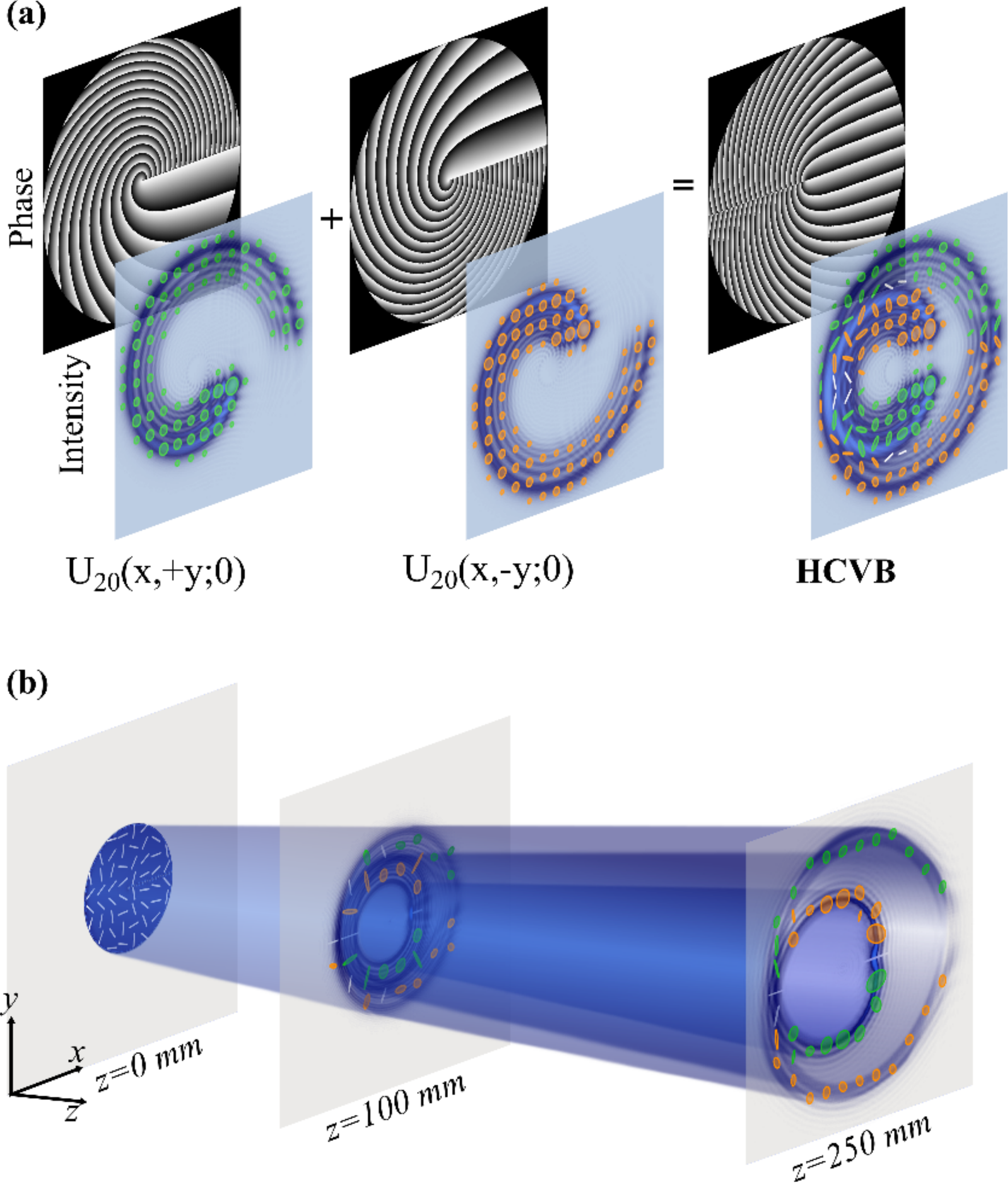}
    \caption{(a) Schematic representation of the superposition of two Helico-Conical beams ($\text{U}_{20} (x, +y; K)$ and $\text{U}_{20} (x, -y; K)$) to generate a HCVB$_{20}$ for $K=0$. The back panels show the phase distribution of each beam (phase of the Stokes field in the case of the HCVB) and the front ones their intensity distribution overlapped with the corresponding polarisation distribution. (b) Conceptual illustration of the propagation-dependent evolution of the HCVB (for this case we used $K=1$). Here, green and orange ellipses represent right- and left-handed elliptical polarisation, respectively, and white segments linear polarisation states.}
    \label{img_conceptHCVB} 
\end{figure}

In general, at a given plane $z$ the transverse dependence of a vector beam $\Vec{E}$ can be written as
\begin{align}
    \Vec{E}(\Vec{r}) = E_L(\Vec{r})\exp[i\phi_L(\Vec{r})]{\bf \hat{e}_L} + E_R(\Vec{r})\exp[i\phi_R(\Vec{r})]{\bf \hat{e}_R},
\end{align}
where ${\bf r}= (x,y) = (r\cos \theta, r \sin\theta)$ are transverse coordinates with $r=\sqrt{x^2 + y^2}$ and $\theta=\arctan{(y/x)}$. Further, ${\bf \hat{e}_R}$ and ${\bf \hat{e}_L}$ represent the unitary vectors associated to right and left circular polarisation, respectively.  In particular, we are interested in the HCVB expressed as
\begin{multline}\label{Eq_VB}
    {\bf HCVB}_\ell({\bf r}, z=0;K) = \cos \beta \:\text{U}_{\ell} (x, \pm y; K) {\bf \hat{e}_R} \\
+ \exp(i\alpha)\sin\beta \:\text{U}_{\ell} (x, \mp y; K){\bf \hat{e}_L},    
\end{multline}
where $\beta$ is a weighting factor, $\alpha$ is an intermodal phase, and $\text{U}_\ell({\bf r}; K) = \text{circ}(r/r_0)\exp[i\phi_\ell({\bf r}; K)]$ represents the HCB endowed with the characteristic helico-conical phase,
\begin{equation}
    \phi_\ell({\bf r};K)=\ell\theta\left(K-\frac{r}{r_0}\right),
    \label{ec_orthogonalPhases}
\end{equation}
and the characteristic constant $K$ that takes values $0$ or $1$, whit $r_0$ a normalisation factor of the radial coordinate. Additionally, $\text{circ}(\cdot)$ is the circle function, and $\ell\in\mathbb{Z}$ is the topological charge of the beam.  Of great relevance is the sign of the $y$ coordinate in $\text{U}_{\ell} (x, \pm y; K)$, as it indicates the direction in which the phase $ \phi_\ell({\bf r};K)$ increases, clockwise for $+y$ and anticlockwise for $-y$. Furthermore, the study of the propagation properties of HCVB, can be achieved via the well--known Kirchhoff's diffraction integral \cite{Goodman1996}. Figure \ref{img_conceptHCVB} (a) illustrates schematically the generation process of HCVB from the superposition of two HCB with orthogonal circular polarisation, for the case $K=1$ and at the propagation distance $z=120$ mm. From left to right we show the phase (back panels) and intensity distribution overlapped with polarisation (front panels) of the beams $\text{U}_{20}(x,+y; 0)$ and $\text{U}_{20}(x,-y; 0)$, carrying right- and left-handed circular polarisation, represented by green and orange ellipses, respectively. The last two panels show the generated vector beam, the phase of the Stokes field (computed from $S = S_1 + iS_2$) in the back and intensity-polarisation distribution in the front. In addition, Fig. \ref{img_conceptHCVB} (b) shows a schematic representation of the propagation-dependent separation of the generated HCVB$_{20}$ for $K=0$, showing the intensity and transverse polarisation distribution at three different planes $z=0$ mm, $z=100$ mm and $z\to\infty$, showcasing the separation of both spatial modes.

\begin{figure}[tb]
    \centering    \includegraphics[width=0.48\textwidth]{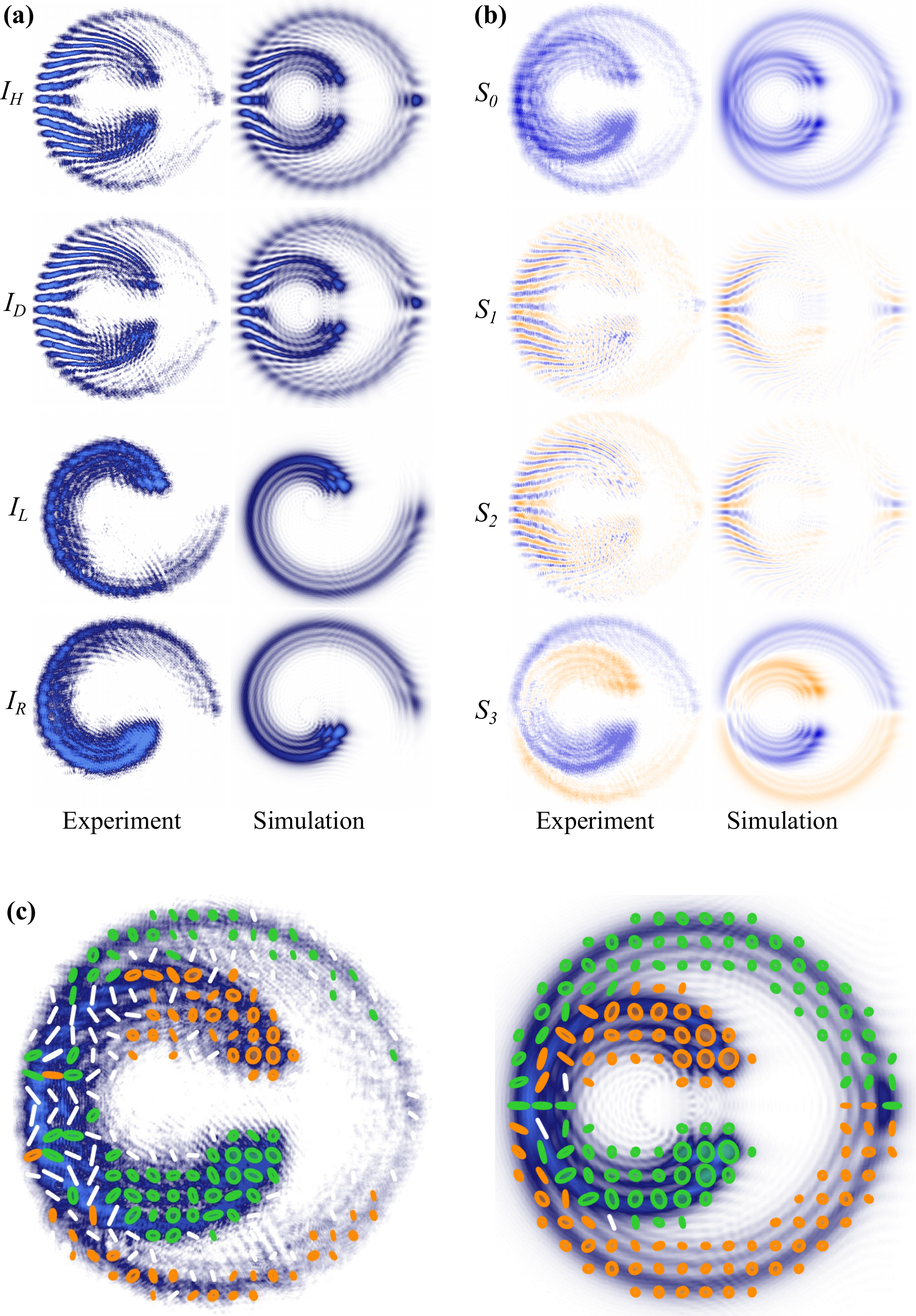}
    \caption{Example polarization reconstruction a HCVB with $K=0$ at a propagation distance of $z=180$ mm. (a) Intensity measurements required to compute the Stokes parameters shown in (b). (c) Intensity distribution overlapped with polarisation distribution reconstructed from the Stokes parameters. In all cases, the experimental results are shown on the left, and numerical simulations on the right.}
    \label{img_stokesPolarimetry} 
\end{figure}

To generate experimentally the HCVB, we implemented an optical setup based on a reflective liquid crystal Spatial Light Modulator (SLM) from Holoeye and a common path interferometer, as detailed in \cite{Perez-Garcia2017}. The polarisation distribution along the transverse plane of the generated vector beam is reconstructed using spatially-resolved Stokes polarimetry. To this end we calculated the Stokes parameters from four intensity measurements as \cite{Goldstein2011}
\begin{equation}\label{Eq.SimplyStokes}
\begin{split}
\centering
 &S_{0}={I_H}+{I_V},\hspace{11.5mm} S_{1}=2{I_H}-S_{0},\hspace{1mm}\\
 &S_{2}=2{I_D}-S_{0},\hspace{10mm} S_{3}=2{I_R}-S_{0},
\end{split}
\end{equation}
where, ${I_H}$, ${I_D}$, ${I_R}$ and ${I_L}$ are the horizontal, diagonal, right- and left-circular polarisation components of the vector beam, respectively, which are measured through a series of polarisation filters, as detailed in \cite{rosales2021Mathieu}. By way of example, in Fig. \ref{img_stokesPolarimetry}(a) we show the corresponding intensities of the HCVB generated with $\ell=20$ and $K=0$ at a propagation distance $z=180$ mm. From these images, the Stokes parameters shown in Fig. \ref{img_stokesPolarimetry}(b) are computed according to Eq. \ref{Eq.SimplyStokes} and from these, the transverse polarisation distribution shown in Fig. \ref{img_stokesPolarimetry}(c) can be reconstructed, as explained in \cite{Goldstein2011}. Here, linear polarisation states are represented by white lines, whereas right- and left-handed polarisation states are represented by green and orange ellipses, respectively. In all cases, experimental results are shown on the left panels, whereas numerical simulations are on the right.

\begin{figure}[tb]
    \centering    \includegraphics[width=0.48\textwidth]{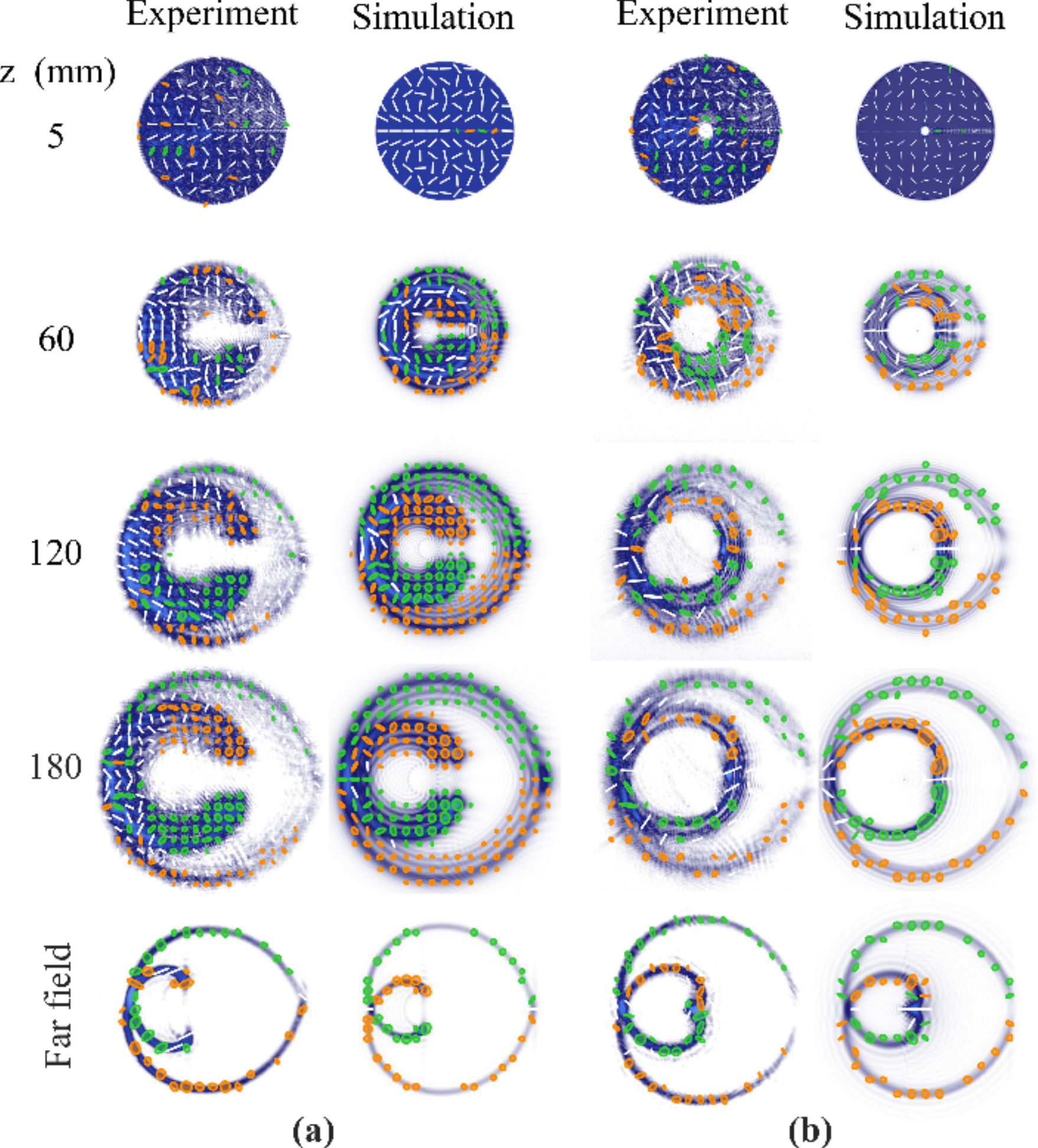}
    \caption{Intensity distribution overlapped with the polarisation distribution at various propagation distances for (a) $K=0$ and (b) $K=1$. In both cases, the experimental results (left) are compared to numerical simulations (right).}
    \label{K=0}
\end{figure}

The propagation-dependent behaviour of HCVB can be observed in a qualitative way from the transverse polarisation distribution. As such, we reconstructed these at various propagation planes, using Stokes polarimetry, for the specific case $\ell=20$. Representative examples are shown in Figs. \ref{K=0} (a) and \ref{K=0} (b) for $K=0$ and $K=1$, respectively, for planes $z=5$ mm, $60$ mm, $120$ mm, $180$ mm and $z\rightarrow\infty$, experimental results on the left panels and numerical simulations obtained through the Kirchhoff diffraction integral on the right \cite{Goodman1996}.  Here, the intensity distribution evolves from one single beam in the near-field to two beams with the shape of spirals at the far-field. In a similar way, the polarisation distribution evolves from a maximally mixed state (at $z=0$), where a mixture of linear polarisation states with different orientations are observed, to a minimum coupling (as $z\to\infty$), where mainly right- and left-handed elliptical polarisation states are present. This qualitative analysis clearly shows a decoupling of both degrees of freedom as a function of propagation.

A more quantitative analysis of the propagation-dependent evolution of HCVB can be done through a recently introduced technique, which relies on the Hellinger Distance \cite{Aiello2022}. This metric suits very well this type of vector modes because it effectively quantifies the local path-like separability of the individual scalar beams constituting the HCVB under analysis. Mathematically, the Hellinger distance is given by, 
\begin{equation}\label{hd}
H(z) =  \;  \sqrt{1 -   \int_{\mathbb{R}^2} \left| \varphi({\bf r},z) \right|  \left| \psi({\bf r},z) \right| \, \text{dA}},  \\[8pt]
\end{equation}
where the functions $\varphi({\bf r},z)$ and $\psi({\bf r},z)$ are precisely the constituting scalar beams of Eq. \ref{Eq_VB} (in this study), which carry orthogonal circular polarisation. The Hellinger distance takes values in the interval [0, 1], 0 when both beams are spatially overlapping each other and 1 when they are completely separated from each other.  

\begin{figure}[tbh]
    \centering
    \includegraphics[width=0.46\textwidth]{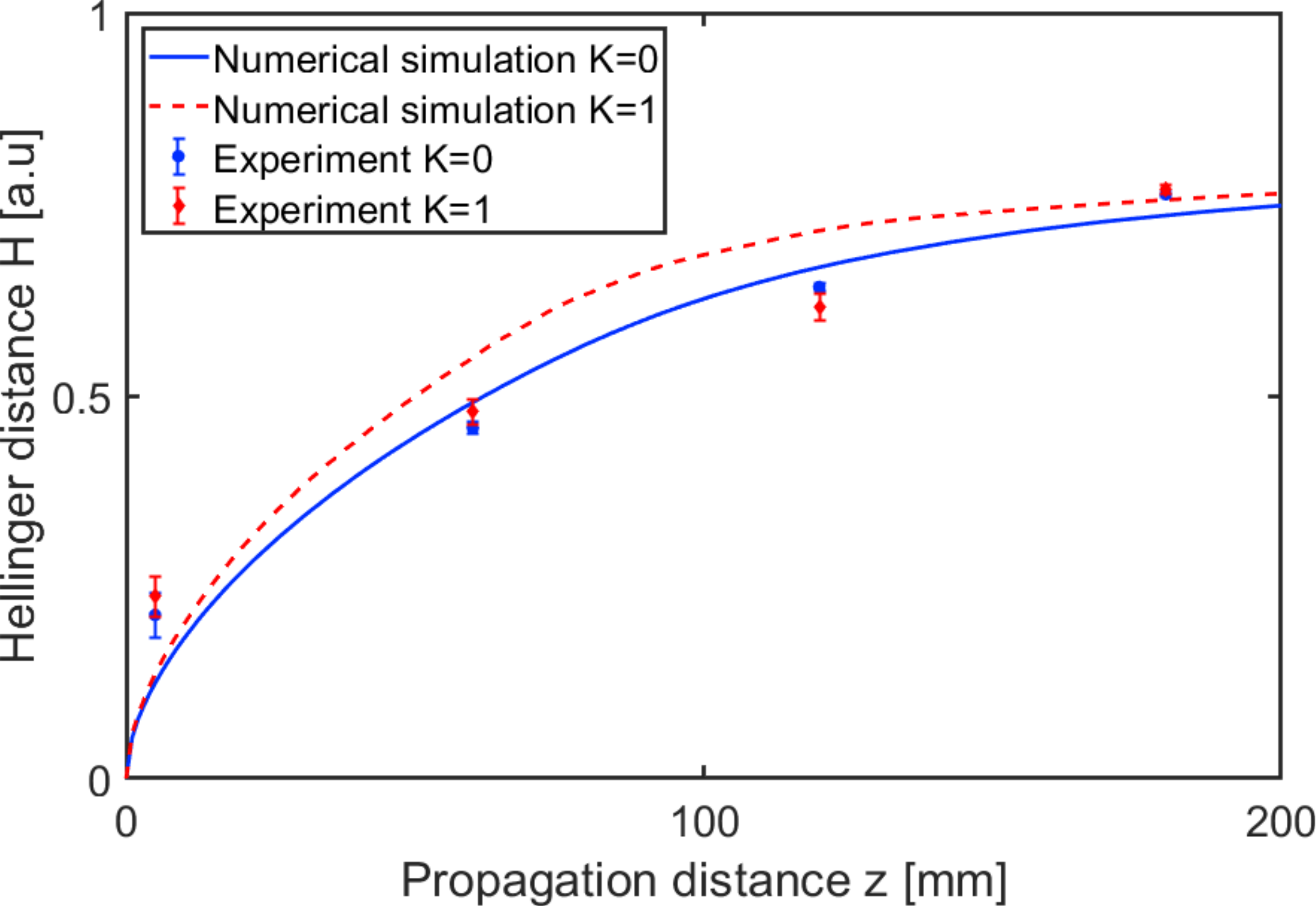}
    \caption{Hellinger distance as a function of the propagation distance for $K=0$ and $K=1$. The continuous curves show results obtained through numerical simulations and the dots correspond to experimental results obtained at $z=5, 60, 120 $, and 180 mm. In the far field, not shown here, $H$ approached the limiting value of $0.91$ for $K=0$ and $0.88$ for $K=1$.}    \label{hellinger}
\end{figure}

Experimentally, $|\varphi({\bf r},z)|$ and $|\psi({\bf r},z)|$ are related to the intensity of each circular polarisation component, at a given plane $z$ and therefore they can be obtained directly from the intensity images used to compute the Stokes parameters. A plot of $H(z)$ as a function of the distance $z$ is shown in Fig. \ref{hellinger} for both cases, $K=0$ and $K=1$. Here, numerical simulations  $K=0$ and $K=1$ are represented as a continuous blue and a dashed red line, respectively and experimental results as blue and red data points. The Hellinger distance effectively describes the evolution of the HCVB, from completely mixed at $z=0$ and for which $H(z)=0$, to quasi--unmixed as $z\to\infty$, for which $H(z)$ tends to the maximum value of 0.91 for the first case and  0.88 for the second. Notice that in the far field, the constituting scalar modes show a small overlap (see Fig. \ref{K=0}) and therefore $H()z$ does not reach the maximum value $H(z)=1$.

In short, in this manuscript we demonstrated a novel class of vector beams obtained through the superposition of circular polarisation states and Helico-Conical Beams, which we termed Helico-Conical Vector beams (HCVB). These beams undergo a propagation-dependent transformation of both, their polarisation and intensity distributions, evolving from a maximum coupling between the spatial and polarisation degrees of freedom to a minimum. In the first, obtained at $z=0$, only linear polarisation states oriented at different angles are observed, whereas in the second, obtained as $z\to\infty$, a quasi-homogeneous distribution of polarisation containing mainly right- and left-handed polarisation states is observed. This behaviour was qualitatively investigated by reconstructing the transverse polarisation distribution at various planes, using Stokes polarimetry, and quantified through the Hellinger distance, a recently proposed measure that accurately describes the degree of nonseparability of spatially-disjoint vector beams. It is wort emphasising that even though here we only presented the case of HCVB with the same topological charge $\ell$, in the superposition given in Eq.~\ref{Eq_VB}, the concepts outlined in this work can be extended, for example to different values of $\ell$ and helicities, or to other types of Helico-conical beams. In addition, the rich polarisation structures that we observed in this study is just an example of the many more that can be obtained and calls for further research. Finally, we anticipate that such vector modes will find applications in optical tweezers, for instance, to sort chiral microparticles \cite{Li2019Sorting} as well as in information encryption. 

%\section*{Funding}
%This research was supported by Zhejiang Provincial Natural Science Foundation of China under Grant No. LQ23A040012, Science Foundation of Zhejiang Sci-Tech University (ZSTU) under Grant No. 22062025-Y and the National Natural Science Foundation of China (61975047).

%\section*{Disclosures}
%The authors declare that there are no conflicts of interest related to this article.

%\bibliography{references}
%\bibliographyfullrefs{references}
\end{document}